\global\def\draftcontrol{0}

   \def\versionno{ 12bps -- draft   }

\catcode`\@=11

\expandafter\ifx\csname draftcontrol\endcsname\relax\global\def\draftcontrol{0}
\fi

{\count255=\time\divide\count255 by 60
\xdef\hourmin{\number\count255}
\multiply\count255 by-60\advance\count255 by\time
\xdef\hourmin{\hourmin:\ifnum\count255<10 0\fi\the\count255}}
\def\draftdate{\number\month/\number\day/\number\year\ \ \ \hourmin }

\newcommand\makepapertitle{\par
  \begingroup
    \renewcommand\thefootnote{\@fnsymbol\c@footnote}%
    \def\@makefnmark{\rlap{\@textsuperscript{\normalfont\@thefnmark}}}%
    \long\def\@makefntext##1{\parindent 1em\noindent
            \hb@xt@1.8em{%
                \hss\@textsuperscript{\normalfont\@thefnmark}}##1}%
     \newpage
     \global\@topnum\z@   
     \@makepapertitle
     \thispagestyle{empty}\@thanks
  \endgroup
  \setcounter{footnote}{0}%
  \global\let\thanks\relax
  \global\let\makepapertitle\relax
  \global\let\@makepapertitle\relax
  \global\let\@thanks\@empty
  \global\let\@author\@empty
  \global\let\@date\@empty
  \global\let\@title\@empty
  \global\let\title\relax
  \global\let\author\relax
  \global\let\date\relax
  \global\let\and\relax
  \def\version{\let\version\@version\@gobble}
}
\def\@makepapertitle{%
  \newpage
   \ifnum\draftcontrol=1 {}
   \version\versionno
   \vskip 3em%
   \else
   \hfill\hbox to 3cm {\parbox{4cm}{\@pubnum}\hss}%
   \vskip 3em%
   \fi
   \begin{center}%
   \let \footnote \thanks
     {\LARGE {\@title}}%
     \vskip 1.5em%
     {\normalsize
       \lineskip .5em%
       \begin{tabular}[t]{c}%
         \@author
       \end{tabular}\par}%
     \vskip 1.5em%
     {\@bstract}%
     \end{center}%
     \vskip 1.5em
     \@date%
   \par
}

\gdef\@pubnum{}
\def\pubnum#1{%
  \gdef\@pubnum{#1}}

\gdef\@bstract{}
\def\Abstract#1{%
  \gdef\@bstract{%
   \parbox{\textwidth-0pc}{%
   \centerline{\bf Abstract}\penalty1000%
\kern.2cm%
\noindent
\renewcommand\baselinestretch{1.0}%
{#1}}}
}

\def\ps@paper{\let\@mkboth\@gobbletwo%
     \ifnum\draftcontrol=1
	\def\@oddfoot{\hbox to \textwidth{\tiny \versionno \hfil\tiny\draftdate}%
	\hskip -\textwidth \hbox to \textwidth{\hfil\rm\thepage\hfil}}%
     \else\def\@oddfoot{\hbox to \textwidth{\hfil\rm\thepage\hfil}}
     \fi
     \let\@evenfoot\@oddfoot
}

\def\body{\clearpage
          \pagestyle{paper}
	}

\def\@version#1{\ifnum\draftcontrol=1
\typeout{}\typeout{#1}\typeout{}
\vskip3mm\centerline{\hbox{\fbox{\normalsize{\tt DRAFT -- #1 -- }
                   {\draftdate}}}}\vskip3mm
\fi}
\let\version\@version
\long\def\eqlabel#1{\ifnum\draftcontrol=1
                    \tag@false  
                    \tag*{(\theequation) \hbox to -0.2cm{\hspace{0cm}\small{#1}\hss}}
                    \refstepcounter{equation}
                    \edef\@currentlabel{\theequation}
                    \ltx@label{#1}          
                    \else
                    \label{#1}
                    \fi
                    }
\let\st@bibitem\@bibitem
\let\st@lbibitem\@lbibitem
\ifnum\draftcontrol=1
  \def\@bibitem#1{%
    \st@bibitem{#1}\a@@label{#1}\ignorespaces}
  \def\@lbibitem[#1]#2{%
    \st@lbibitem[#1]{#2}\a@@label{#2}\ignorespaces}
  \def\a@@label#1{%
    \gdef\a@lab{\smash{\normalfont\small#1}}
    \ifvmode
      \if@inlabel
        \global\setbox\@labels\hbox{%
          \llap{\a@lab\let\a@lab\relax
                \kern\@totalleftmargin\kern\marginparsep}%
          \box\@labels}%
      \fi
    \fi}
\fi

\documentclass[12pt,letterpaper]{article}

\usepackage{amsmath,amssymb,array,calc,rotating,epsfig,psfrag}
\usepackage[nosort]{cite}

\ifnum\draftcontrol=1
\fi

\renewcommand\baselinestretch{1.25}
\renewcommand\section{\@startsection {section}{1}{\z@}%
                                   {-3.5ex \@plus -1ex \@minus -.2ex}%
                                   {2.3ex \@plus.2ex}%
                                   {\normalfont\large\bfseries}}
\renewcommand\subsection{\@startsection{subsection}{2}{\z@}%
                                   {-3.25ex\@plus -1ex \@minus -.2ex}%
                                   {1.5ex \@plus .2ex}%
                                   {\normalfont\normalsize\bfseries}}
\renewcommand\subsubsection{\@startsection{subsubsection}{3}{\z@}%
                                   {-3.25ex\@plus -1ex \@minus -.2ex}%
                                   {1.5ex \@plus .2ex}%
                                   {\normalfont\normalsize\it}}
\renewcommand\paragraph{\@startsection{paragraph}{4}{\z@}%
                                   {-3.25ex\@plus -1ex \@minus -.2ex}%
                                   {1.5ex \@plus .2ex}%
                                   {\normalfont\normalsize\bf}}

\numberwithin{equation}{section}

\def\ie{{\it i.e.}}

\def\revise#1       {\raisebox{-0em}{\rule{3pt}{1em}}%
                     \marginpar{\raisebox{.5em}{\vrule width3pt\
                     \vrule width0pt height 0pt depth0.5em
                     \hbox to 0cm{\hspace{0cm}{%
                     \parbox[t]{4em}{\raggedright\footnotesize{#1}}}\hss}}}}

\def\cala         {{\cal A}}

\def\cald         {{\cal D}}

\def\caln         {{\cal N}}
\def\calo         {{\cal O}}

\def\del          {\partial}

\def\sqr#1#2{{\vcenter{\vbox{\hrule height.#2pt
 \hbox{\vrule width.#2pt height#1pt \kern#1pt
 \vrule width.#2pt}\hrule height.#2pt}}}}

\newcommand{\ft}[2]{{\textstyle{\frac{#1}{#2}}}}

\def\a{\alpha}
\def\e{\epsilon}
\def\r{\rho}
\def\t{\theta}
\def\g{\gamma}

\catcode`\@=12

\begin{document}


\title{Coarse-graining 1/2 BPS geometries of type IIB supergravity}

\pubnum{%
hep-ph/0409271}
\date{September 2004}

\author{
Alex Buchel\\[0.4cm]
\it Perimeter Institute for Theoretical Physics\\
\it Waterloo, Ontario N2J 2W9, Canada\\[0.2cm]
\it Department of Applied Mathematics\\
\it University of Western Ontario\\
\it London, Ontario N6A 5B7, Canada\\
}

\Abstract{
Recently Lin, Lunin and Maldacena (LLM) \cite{llm} explicitly mapped
$1/2$ BPS excitations of type IIB supergravity on $AdS_5\times S^5$
into free fermion configurations. We discuss thermal coarse-gaining of
LLM geometries by explicitly mapping the corresponding equilibrium
finite temperature fermion configuration into supergravity. Following
Mathur conjecture, a prescription of this sort should generate a
horizon in the geometry. We did not find a horizon in finite
temperature equilibrium LLM geometry. This most likely is due to the
fact that coarse-graining is performed only in a half-BPS sector of
the full Hilbert space of type IIB supergravity.  For temperatures
much less than the $AdS$ curvature scale the equilibrium background
corresponding to nearly degenerate dual fermi-gas is found
analytically.        
}


\makepapertitle

\body

\version\versionno

\section{Introduction}
In recent years motivated
and guided by the AdS/CFT correspondence \cite{m9711},  new lessons have been learned
about time-like singularities and their resolution in supergravity and string
theory. Firstly, the goal of exhibiting new tests of the duality has led to 
the construction of many new exact supergravity backgrounds. This includes
metrics on non-compact Calabi-Yau spaces with and without fluxes, 
$G_2$-holonomy metrics, etc. Secondly, the correspondence with gauge theory 
has hinted at the correct interpretation and resolution of singularities.
The mechanisms include confinement, chiral symmetry breaking, generation of a
mass gap, etc. These recent advances have complemented the existing knowledge
about singularities from closed strings and D-branes. 
On the contrary, a space-like singularity, such as  behind the 
horizon of the Schwarzschild black hole or the Big Bang singularity in cosmology, 
is poorly understood. 

An interesting approach towards understanding  space-like
singularities behind  black hole horizons is initiate by Mathur and
collaborators \cite{m1,m2,m3,m4,m5}.  The implication of the proposal
is that these singularities are not physical: the long-suspected
picture of a 'stringy' Schwarzschild horizon with a smooth region inside
it\footnote{Having {\it only} the local large curvature region of the
geometry  resolved by the string corrections.} is just an
artifact of a classical analysis. Rather, the correct physical picture
is akin the one arising in statistical mechanics. Much like equilibrium 
thermodynamics arises from the thermal coarse-graining over an ensemble of 
microstates of a dynamical system, the black hole solution (with a Schwarzschild
horizon) is a result of a thermal averaging over its true microstate-geometries.
These microstate-geometries are precisely the ones that the Bekenstein-Hawking 
entropy accounts for. Each one of the black hole microstates represents a smooth, 
singularity and horizon-free geometry. Currently the detailed understanding of black hole microstates 
exists for a rather degenerate case: a supersymmetries D1-D5 system in Ramond ground state. 
This black hole has zero temperature and zero macroscopic area of the horizon. 
In the last couple months  very encouraging results appear \cite{r1,r2,r3,r4,r5,r6,r7} 
that suggest that  Mathur's program of explaining the microstates of a black hole
can be extended to a three-charge brane configuration, which, though supersymmetric,
has a finite horizon area.  
         
Understanding the microstates of a non-extremal black hole is a daunting task. For one reason,
it is not even clear that all such states allow for a gravitational description\footnote{Some microstates
might not be even geometrical.}. But even if identified, a prescription has to be formulated 
how to coarse-grain these micro-geometries. Such a prescription is evident in case there is a 
dual, purely field theoretical, formulation of geometries: all we have to do it to form a canonical
(or grand canonical) ensemble of the dual field theory and map resulting thermal equilibrium configuration 
into the gravity. It is natural to conjecture that this procedure would produce a geometry 
with a Schwarzschild horizon. We would like to stress that above conjecture (as well as  Mathur's interpretation of a 
black hole) is a very intuitive extension of the finite temperature gauge theory/string theory 
correspondence \cite{gg} where (in the simplest case) a large Schwarzschild black hole in global 
$AdS_5$ geometry corresponds to equilibrium finite temperature $\caln=4$ $SU(N)$ supersymmetric 
Yang-Mills (SYM) theory on $S^3\times R$. 
    
In \cite{llm} (LLM) the authors explicitly constructed smooth (and without horizon)
type IIB supergravity backgrounds holographically dual to chiral primary operators of 
$\caln=4$ SYM. These are operators with conformal weight $\triangle=J$, where $J$ 
is a particular $U(1)$ charge of the R-symmetry group. The LLM geometries are a subset of
micro-geometries one would have to use to realize a non-extremal black hole solution\footnote{This black hole 
must have a chemical potential conjugate to the $U(1)$ R-symmetry charge.
Thermodynamics of such a  black hole in relation to a string dual of a large R-charge 
sector of SYM was discussed in \cite{bp}. } with  
$AdS_5$ asymptotic. In this paper we discuss thermal coarse-graining of LLM geometries.
This is achieved by mapping the finite temperature 
free fermion description \cite{b,llm}  of $1/2$ BPS states of \cite{llm} into supergravity. 
The resulting thermally equilibrium supergravity background does not have a horizon. 
This is perhaps not  surprising: one expects to recover the AdS-Schwarzschild solution 
upon thermal averaging over the {\it whole} Hilbert space of supergravity (string?) excitations 
in $AdS_5\times S^5$ rather than just over its $1/2$ BPS sector.    
Nonetheless, it is interesting to explore if supergravity backgrounds obtained in such a 
coarse-graining exhibit thermodynamic properties. We discuss this in a future publication. 

The paper is organized as follows. In the next section we briefly review the LLM geometries. 
Using the free fermion description of LLM supergravity states we explain how to implement 
their finite temperature coarse-graining. In the limit of nearly degenerate fermi gas the 
dual geometry can be found analytically. We conclude in section 3.

\section{LLM 1/2 BPS geometries at finite temperature}
Type IIB supergravity backgrounds dual to $1/2$ BPS states
of $\caln=4$ $SU(N)$ SYM on $S^3\times R$ \cite{llm} are  characterized by a single 
function in $R^2:\ \{x_1,x_2\}$
\begin{equation}
\begin{split}
z\equiv z(x_1,x_2,0)=&1\,,\qquad \{x_1,x_2\}\in \cald\subset R^2\,,\\
z=&0\,,\qquad {\rm otherwise}\,.
\end{split}
\eqlabel{zdef}
\end{equation}  
The $R^2$ in question has an interpretation as a phase space of (dual) free fermions \cite{b}, and $\cald$ is a 
domain in this phase space occupied by a fermion droplet realizing a given chiral primary of SYM.
The area $\cala(\cald)$ of a droplet is quantized in units of 
\begin{equation}
\hbar\equiv 2\pi \ell_p^4=2\pi g_s\left(\a'\right)^2\,,
\eqlabel{hb}
\end{equation} 
so that 
\begin{equation}
\frac{\cala(\cald)}{2\pi \hbar}=N
\eqlabel{n}
\end{equation}
is the total 5-form flux of the supergravity background.
The quantum energy\footnote{In units of $1/R_{AdS_5}$.} 
of a fermion at $(x_1,x_2)$  in the phase space
is 
\begin{equation}
\e\equiv \e(x_1,x_2)=\frac{x^2_1+x_2^2}{2\hbar}\equiv \frac{r^2}{2\hbar}\,.
\eqlabel{en}
\end{equation} 
A Fermi surface of zero temperature fermion configuration is a circle of radius 
\begin{equation}
r_0^{(T=0)}=(2\hbar N)^{1/2}\,,
\eqlabel{r0}
\end{equation} 
about the origin in the phase space. The energy of a particular  fermion droplet $\cald$
above the energy of the ground state is identified with the conformal dimension $\triangle$ of the SYM 
chiral primary, or an ADM mass of the corresponding supergravity background
\begin{equation}
\begin{split}
\triangle=J=&\frac{1}{16\pi^3\ell_p^8}\left[\int_{\cald}d^2x\left(x_1^2+x_2^2\right)-\frac{1}{2\pi}
\left(\int_{\cald}d^2x\right)^2\right]\\
=&\int_{\cald}\frac{d^2x}{2\pi\hbar}\ \frac{\ft 12 (x_1^2+x_2^2)}{\hbar}-\frac 12 \left(\int_{\cald}\frac{d^2x}{2\pi\hbar}\right)^2\,.
\end{split}
\eqlabel{energy}
\end{equation}
Thus,  zero temperature configuration \eqref{r0} corresponds to $\triangle=0$, \ie, 'vacuum' (or $AdS_5\times S^5$) configuration
\begin{equation}
\triangle^{(T=0)}=\int_0^{r_0}\frac{2\pi r dr}{2\pi\hbar}\ \frac{r^2}{2\hbar}-\frac 12 \left(\frac{\pi r_0^2}
{2\pi\hbar}\right)^2 =0\,.
\eqlabel{vauum}
\end{equation} 

Given $z(x_1,x_2,0)$ of \eqref{zdef} the corresponding supergravity background is \cite{llm}
\begin{equation}
\begin{split}
ds^2=&-h^{-2}\left(dt+V_i dx^i\right)^2+h^2\left(dy^2+dx^idx^i\right)+y e^G\ d\Omega_3^2+ye^{-G}d\tilde\Omega_3^2\,,\\
h^{-2}=&2y\cosh G\,,\\
y\del_y V_i=&\e_{ij}\del_j z\,,\qquad y\del_i V_j=\e_{ij}\del_y z\,,\\
z=&\frac 12 \tanh G\,,\\
F_{(5)}=&F_{\mu\nu} dx^{\mu}\wedge dx^{\nu}\wedge d\Omega_3+\tilde{F}_{\mu\nu} dx^{\mu}\wedge dx^{\nu}\wedge 
d\tilde{\Omega}_3\,,\\
F=&dB_t\wedge \left(dt+V\right)+B_tdV+d\hat{B}\,,\\
\tilde{F}=&d\tilde{B}_t\wedge \left(dt+V\right)+\tilde{B_t}dV+d\tilde{\hat{B}}\,,\\
B_t=&-\frac 14 y^2 e^{2G}\,,\qquad \tilde{B}_t=-\frac 14 y^2e^{-2G}\,,\\
d\hat{B}=&-\frac 14 y^3\ \star_3 d\left(\frac{z+\ft 12}{y^2}\right)\,,\qquad 
d\tilde{\hat{B}}=-\frac 14 y^3\ \star_3 d\left(\frac{z-\ft 12}{y^2}\right)\,,
\end{split}
\eqlabel{equations}
\end{equation}
where $i=1,2$ and $\star_3$ is the flat space Hodge dual in the three dimensions parameterized by $y, x_1, x_2$.
Also
\begin{equation}
\begin{split}
z\equiv z(x_1,x_2,y)=&\frac{y^2}{\pi}\int_{R^2}\frac{z(x_1',x_2',0)dx_1'dx_2'}{[({\bf{x-x'}})^2+y^2]^2}=
\frac{y^2}{\pi}\int_{\cald}\frac{dx_1'dx_2'}{[({\bf{x-x'}})^2+y^2]^2}\,,\\
V_i\equiv V_i(x_1,x_2,y)=&\frac{\e_{ij}}{\pi}\int_{R^2}\frac{z(x_1',x_2',0)(x_j-x_j')dx_1'dx_2'}
{[({\bf{x-x'}})^2+y^2]^2}=\frac{\e_{ij}}{\pi}\int_{\cald}\frac{(x_j-x_j')dx_1'dx_2'}
{[({\bf{x-x'}})^2+y^2]^2}\,.
\end{split}
\eqlabel{zvdef}
\end{equation}
The 3-form fluxes and  an axion-dilaton vanish.  

Generically, a supergravity background \eqref{equations} is time-dependent. A static 
solution arises whenever geometry has 
a second Killing vector. This is the case if $\cald$ is rotationally invariant about the origin in $R^2$, as for the circular 
fermionic droplets, or a fermion configurations corresponding to concentric circles. In the former case (say 
zero temperature fermion configuration) an explicit solution is \cite{llm}
\begin{equation}
\begin{split}
z(r,y;r_0)=&\frac{r^2-r_0^2+y^2}{2\sqrt{(r^2+r_0^2+y^2)-4r^2r_0^2}}\,,\\
V_{\phi}=&-\frac 12 \left(\frac{r^2+y^2+r_0^2}{\sqrt{(r^2+r_0^2+y^2)^2-4r^2r_0^2}}-1\right)\,,
\end{split}
\eqlabel{expsol}
\end{equation}
where $r_0$ is the radius of the fermion droplet, and $\{r,\phi\}$ are the polar coordinates in $x_1, x_2$ plane. 
With \eqref{expsol} and the change of variables \cite{llm}
\begin{equation}
\begin{split}
y=&r_0\sinh\r\sin\t\,,\\
r=&r_0\cosh\r\cos\t\,,\\
\psi=&\phi+t\,,
\end{split}
\eqlabel{cv}
\end{equation}
one recovers from \eqref{equations} the standard $AdS_5\times S^5$ metric
\begin{equation}
ds^2=r_0\left[-\cosh^2\r dt^2+\sinh^2\r d\Omega_3^2+d\r^2+d\t^2+\cos^2\t d\psi^2+\sin^2\t d\tilde{\Omega}_3^2\right]\,.
\eqlabel{ads5}
\end{equation}

We now turn to thermal coarse-graining of LLM geometries. We propose that the equilibrium geometry 
corresponding to coarse-graining $1/2$ BPS states of LLM at temperature\footnote{In units of $1/R_{AdS_5}$.} $T$
is an LLM supergravity background with the source $z(x_1,x_2,0)$ in \eqref{zdef} identical to the 
configuration of the free fermion gas in the phase space $R^2$ at temperature $T$.
As such, the resulting coarse-grained state is supersymmetric. 
At first sight  this might look counterintuitive: how can supersymmetry be consistent 
with finite temperature? The resolution is simply that this is an artifact of the 
thermal averaging in $1/2$ BPS sector of the Hilbert 
space: if {\it all} states in (grand)canonical ensemble are supersymmetric, 
the thermal state is supersymmetric as well. With above conjecture, it is straightforward to 
obtain the required geometry. We expect the resulting fermion equilibrium configuration $\cald$ to be rotationally 
invariant, \ie, it is (at most) a set of concentric circles.   

The number density of fermions in a phase space at temperature $T$ is given by the Fermi distribution
\begin{equation}
dN=n(r)\ \frac{2\pi r  dr}{2\pi \hbar} \equiv  \frac{1}{e^{(\e(r)-\mu)/T}+1}\ \frac{2\pi r dr}{2\pi \hbar}\,, 
\eqlabel{nr}
\end{equation}  
where $\mu$ is a chemical potential\footnote{Not to be confused with the 
chemical potential conjugate to the $U(1)$ charge $J$.}, and  a fermion energy is given by \eqref{en}.
Keeping the total number $N$ of fermions (or the 5-form flux) fixed, determines the chemical potential
\begin{equation}
N=\int dN=\int_0^\infty\frac{1}{e^{(\e(r)-\mu)/T}+1}\ \frac{2\pi r dr}{2\pi \hbar} =T\ln\left(1+e^{\mu/T}\right)\,,
\eqlabel{mudef}
\end{equation} 
or 
\begin{equation}
\mu=T\ln\left(e^{N/T}-1\right)\,.
\eqlabel{mucom}
\end{equation}
Notice that at small temperatures (degenerate fermi gas)
\begin{equation}
\mu\approx N\,,\qquad T\ll N\,.
\eqlabel{mudeg}
\end{equation} 
The phase space configuration of the finite temperature fermion gas is a set of concentric rings about the 
origin of the phase space with the 'width' of a ring of radius $r_i$ being equal $n(r_i)\triangle r_i$. 
The radial separation between two consecutive rings is $(1-n(r_i))\triangle r_i$. These leads\footnote{Concentric configurations 
are also discussed in \cite{llm}.} to the 
coarse-grained source function $z^T(r,y)$ as
\begin{equation}
\begin{split}
z^T(r,y)=&\sum_i\ \biggl(z(r,y;r_i+n(r_i)\triangle r_i)-z(r,y;r_i)\biggr)\\
\approx &\sum_i \frac{\del z(r,y;r_i)}{\del r_i}\ n(r_i)\triangle r_i\\
=&\int_0^{\infty} {dx} n(x) \frac{\del z(r,y;x)}{\del x}\,,
\end{split}
\eqlabel{f1}
\end{equation}  
where $z(r,y;x)$ is  given by \eqref{expsol}.
Alternatively, 
\begin{equation}
z^T(r,y)=-\int_0^{\infty} dx z(r,y;x) \frac{\del n(x)}{\del x}\,.
\eqlabel{ztal}
\end{equation} 
From \eqref{ztal} we can immediately recover the zero temperature limit. Indeed, at $T=0$, $n(x)$
is a step function. Thus $\del n(x)/\del x$  is $-\delta(x-r_0^{(T=0)})$, which leads to \eqref{expsol}.
We did not succeed in explicitly evaluating \eqref{f1} for arbitrary temperature. 
Since geometry dual to $z^T$ \eqref{f1} fermion configuration 
 is an LLM $1/2$ BPS state, it is singularity-free and does not have a horizon.

In the rest of this section we present coarse-grained  geometry in the limit of 
vanishingly small temperature. The computation is a standard one in the statistical 
physics of the degenerate fermi gas \cite{ll}. Given that for an arbitrary  smooth function 
$f(\e)$ such that $I$ defined as
\begin{equation}
I=\int_0^{\infty}\frac{f(\e)d\e}{e^{(\e-\mu)/T}+1}\,,
\eqlabel{ill}
\end{equation}
converges, we have
\begin{equation}
I=\int_0^\mu f(\e) d\e+\frac{\pi^2}{6} T^2\ f'(\mu)+\calo(T^4)\,,
\eqlabel{iapp}
\end{equation}
and using change of variables\footnote{For concentric ring distributions 
  $d\psi\equiv d\phi-dt\ \frac{V_\phi y^2}{r^2 z^2 e^{2G}-y^2 V_\phi^2}$.} \eqref{cv}
the resulting geometry can be written as 
\begin{equation}
ds^2=-g_{tt} dt^2+g_{\r\r}\left(d\r^2+d\t^2\right)+g_{\psi\psi} d\psi^2+g_{\Omega_3\Omega_3}\ d\Omega_3^2+g_{\tilde{\Omega}_3\tilde{\Omega}_3}\ 
d\tilde{\Omega}_3^2\,,
\eqlabel{geomef}
\end{equation}
where for $\r>0$ and up to $\calo(T^4)$ terms  
\begin{equation}
\begin{split}
g_{tt}=&r_0 \cosh^2\r\biggl(1-\g\\
&\times
\frac{(3\cos^2\t-1)\cosh^4\r+(3\cos^4\t-6\cos^2\t+2)\cosh^2\r+\cos^2\t(2-\cos^2\t)}{(\cosh^2\r-\cos^2\t)^3}
\biggr)\,,\\
g_{\r\r}=&r_0\biggl(1+\g\  (\cosh^2\r+\cos^2\t-2)\\
&\times \frac{(3\cos^2\t-1)\cosh^4\r-\cos^2\t(4-3\cos^2\t)\cosh^2\rho-\cos^4\t}{(\cosh^2\r-\cos^2\t)^4}
\biggl)\,,\\
g_{\psi\psi}=&r_0\cos^2\t\biggl(1-\g\ \frac{1}{(\cosh^2\r-\cos^2\t)^4}\\
&\times \biggl[(3-9\cos^2\t)\cosh^6\r-(6\cos^4\t-19\cos^2\t+4)\cosh^4\r\\
&+\cos^2\t(3\cos^4\t+3\cos^2\t-8)\cosh^2\r-\cos^6\t
\biggr]
\biggr)\,,\\
g_{\Omega_3\Omega_3}=&r_0\sinh^2\r\biggl(1-\g\\
&\times\frac{(3\cos^2\t-1)\cosh^4\r-\cos^2\t(4-3\cos^2\t)\cosh^2\r-\cos^4\t}{(\cosh^2\r-\cos^2\t)^3}
\biggr)\,,\\
g_{\tilde{\Omega}_3\tilde{\Omega}_3}=&r_0\sin^2\t\biggl(1+\g\\
&\times\frac{(3\cos^2\t-1)\cosh^4\r-\cos^2\t(4-3\cos^2\t)\cosh^2\r-\cos^4\t}{(\cosh^2\r-\cos^2\t)^3}
\biggr)\,,
\end{split}
\eqlabel{a1}
\end{equation}
and 
\begin{equation}
d\psi\equiv d\phi+dt\biggl(1-\g\ \frac{2(\cos^2\t+\sinh^2\r)\cosh^2\r}{(\cosh^2\r-\cos^2\t)^3}\biggr)\,.
\eqlabel{aa1}
\end{equation}
In \eqref{a1},\eqref{aa1} $r_0$ is given by \eqref{r0}, also  we introduced 
\begin{equation}
\g\equiv \frac {2 \pi^2 T^2 \hbar^2}{3r_0^4}=\frac{\pi^2 T^2}{6 N^2}\,.
\eqlabel{a2}
\end{equation}

\section{Conclusion}
In this paper we constructed supergravity background dual to thermal averaging 
over LLM geometries. The resulting geometry is singularity free. It does not have a 
horizon either. 
A similar prescription when implemented on the full Hilbert space of 
the type IIB string excitations in asymptotic $AdS_5\times S^5$ background is 
expected to produce an AdS-Schwarzschild black hole. It is interesting to 
explore whether this coarse-grained geometry exhibit thermal properties of the 
dual finite temperature free fermions. One specific question is whether 
the ADM mass of the thermal geometry agrees with the free fermion gas energy.        
In the absence of a horizon, how does the coarse-grained geometry encode the entropy of the 
free fermion gas?

\section*{Acknowledgments}
I would like to thank I.~Bena, E.~Gorbar and  V.~Miransky for helpful conversations. 
Research at  Perimeter Institute is supported in part 
by funds from NSERC of Canada. Additional support by an 
NSERC Discovery grant is gratefully acknowledged.

\end{document}